\documentclass[a4paper,11pt]{article}
\usepackage{pos}
\usepackage{braket}
\usepackage{float}
\usepackage{caption}
\usepackage{subcaption}
\usepackage{url}

\usepackage{xcolor}

\definecolor{darkgreen}{rgb}{0.0, 0.5, 0.0}

\title{Hamiltonian limit of lattice QED in 2+1 dimensions}

\author[a,b]{L.\ Funcke}
\author*[c]{C.\ F.\ Groß}
\author[d]{K.\ Jansen}
\author[d,e]{S.\ Kühn}
\author*[c]{S.\ Romiti}
\author[c]{C.\ Urbach}

\affiliation[a]{Transdisciplinary Research Area ``Building Blocks of Matter and Fundamental Interactions'' (TRA Matter) and Helmholtz Institute for Radiation and Nuclear Physics (HISKP), University of Bonn, Nußallee 14-16, 53115 Bonn, Germany}

\affiliation[b]{Center for Theoretical Physics, Co-Design Center for Quantum Advantage, and NSF AI Institute for Artificial Intelligence and Fundamental Interactions, Massachusetts Institute of Technology, 77 Massachusetts Avenue, Cambridge, MA 02139, USA}

\affiliation[c]{
HISKP and Bethe Center for Theoretical Physics, Rheinische Friedrich-Wilhelms-Universität Bonn, Nußallee 14-16, 53115 Bonn, Germany}

\affiliation[d]{
Deutsches Elektronen-Synchrotron DESY, Platanenallee 6, 15738 Zeuthen, Germany
}

\affiliation[e]{Computation-Based Science and Technology Research Center, The Cyprus Institute, 20 Kavafi Street, 2121 Nicosia, Cyprus}

\emailAdd{lfuncke@uni-bonn.de}
\emailAdd{christiane.gross@uni-bonn.de}
\emailAdd{karl.jansen@desy.de}
\emailAdd{stefan.kuehn@desy.de}
\emailAdd{simone.romiti@uni-bonn.de}
\emailAdd{urbach@hiskp.uni-bonn.de}

\abstract{

The Hamiltonian limit of lattice gauge theories can be found by extrapolating the results of anisotropic lattice computations, i.e., computations using lattice actions with different temporal and spatial lattice spacings ($a_t\neq a_s$), to the limit of $a_t\to 0$.
In this work, we present a study of this Hamiltonian limit for a Euclidean $U(1)$ gauge theory in 2+1 dimensions (QED3), regularized on a toroidal lattice.
The limit is found using the renormalized anisotropy $\xi_R=a_t/a_s$, by sending $\xi_R \to 0$ while keeping the spatial lattice spacing constant.
We compute $\xi_R$ in $3$ different ways: using both the ``normal'' and the ``sideways'' static quark potential, as well as the gradient flow evolution of gauge fields. 
The latter approach will be particularly relevant for future investigations of combining quantum computations with classical Monte Carlo computations, which requires the matching of lattice results obtained in the Hamiltonian and Lagrangian formalisms.\\\\ Preprint number: MIT-CTP/5480
}

\FullConference{
  The 39th International Symposium on Lattice Field Theory, LATTICE2022 \\
  8-13 August 2022 \\
  Hörsaalzentrum Poppelsdorf, Bonn, Germany
}

\usepackage{hyperref}
\begin{document}
\maketitle

\section{Introduction}

The idea of studying gauge theories on a space-time lattice dates back to 1974-75~\cite{PhysRevD.10.2445, PhysRevD.11.395}. 
Since then, most numerical simulations have been performed in the Lagrangian
formalism, using path integral Monte Carlo techniques at imaginary time~\cite{aoki2022flag}. 
However, in recent years, we have been witnessing a rapid development of Hamiltonian-based simulations using, e.g., tensor networks~\cite{Banuls2018a} or quantum computing~\cite{Banuls2020}. 
In particular, the rapid development of quantum technology~\cite{gill2022quantum} may open  a window for phenomenologically relevant Hamiltonian
simulations 
on quantum computers in the future.
Although there is still a long way to go from both the theoretical and technological perspective, first proof-of-concept simulations of lower-dimensional gauge 
theories have already been performed using quantum computers~(see, e.g., Ref.~\cite{klco2021standard} for a recent review).

In the continuum limit, the Lagrangian and Hamiltonian formalisms are equivalent~\cite{RevModPhys.20.367, dirac2005lagrangian}. On the lattice, the Hamiltonian limit is obtained by sending the temporal lattice spacing to zero, $a_t \to 0$, using an \textit{anisotropic} lattice action~\cite{PhysRevD.70.014504, degrand2006lattice, gattringer2009quantum}.
The anisotropy is introduced through an additional parameter in the lattice action, called the \textit{bare anisotropy} $\xi_0$. 
For $\xi_0\neq 1$, this parameter breaks the symmetry between the temporal and spatial contributions. The resulting \textit{renormalized anisotropy} $\xi_R = a_t/a_s$, defined as the ratio between the temporal and spatial lattice spacings, also deviates from 1.
The Hamitonian limit can be found by sending $\xi_R \to 0$, while keeping $a_s$ fixed.

At finite lattice spacing, there is a non-trivial dependence between the bare parameters and the observables in the Lagrangian and Hamiltonian formalisms, and a matching is required.
This can be done in two ways. 
First, at a given spatial lattice spacing $a_s$, one can match the bare parameters $g_i$ to equally many observables $O_i$, either perturbatively or non-perturbatively. 
Second, one can perform a set of Lagrangian simulations at decreasing temporal lattice spacing $a_t$, extrapolating to the Hamiltonian limit ($a_t\to 0$) in parameter space. 

In this work, we explore the latter approach for a $U(1)$ gauge theory in $2+1$ dimensions. 
In the literature, this theory is often referred to as QED 2+1~(see, e.g., Refs.~\cite{PhysRevD.103.096002, zeitlin1997induced, suzuki1983confinement, agarwal2019antiparticles, PhysRevD.68.034504}) or QED3~(see, e.g, Refs.~\cite{coleman1985no, chester2016towards, pennington1991masses, dorey1992qed3, hands2002non, curtis1992dynamical}).
We choose to study this theory due to the recent proposal of Ref.~\cite{clemente2022strategies} to combine quantum computations of QED 2+1 with classical Monte Carlo computations, 
which requires a matching of the lattice results obtained in the Hamiltonian and Lagrangian formalisms. The generalisation of our study to higher dimensions and $SU(N)$ theories could be straight-forwardly done with simple theoretical modifications, but would of course be computationally challenging.
We note that QED 2+1 is not only relevant for condensed matter systems~\cite{kosinski2012qed2+}, but also
exhibits confinement~\cite{athenodorou2019spectrum} and dynamical mass generation~\cite{PhysRevD.33.3704}.
These features make it a toy model for QCD and a benchmark model for the study of lattice gauge theories.
Most crucially, due to its simple structure, QED 2+1 offers the possibility of near-future simulations in the Hamiltonian formalism using quantum hardware (see, e.g., Ref.~\cite{Haase:2020kaj,Paulson:2020zjd,clemente2022strategies}).

An important advantage of Hamiltonian simulations is the absence of specific numerical issues in Monte Carlo simulations, in particular the sign problem \cite{gattringer2016approaches} and the problem of critical slowing down \cite{gattringer2009quantum}.
Therefore, in general, matching the two formalisms allows to span the whole parameter space and to improve the continuum limit by adding more points to the extrapolation.

The rest of the paper is organized as follows. 
In Sec.~\ref{theory.background}, we introduce the lattice action and some theoretical aspects of the Hamiltonian limit.
Section~\ref{sec:xiR.potential} shows how to compute $\xi_R$ using the static potential, and Sec.~\ref{sec:limit.plaquette} the Hamiltonian limit for the plaquette expectation value.
In Sec.~\ref{sec:xiR.glow}, we discuss the calculation of $\xi_R$ using the gradient flow.
In Sec.~\ref{sec:conclusions}, we provide conclusions and an~outlook.

\section{Theoretical background}
\label{theory.background}

In this work, we study a Euclidean $U(1)$ gauge theory on a periodic (2+1)-dimensional lattice of size  $\Lambda=L^2 \times T = (N_s a_s)^2 \times (N_t a_t)$, where
$N_t$ ($N_s$) is the number of lattice sites in the temporal (spatial) direction, separated by a lattice spacing $a_t$ ($a_s$). 
We use the compact formulation~\cite{romiti2021neutron, Marco.Panero.2005} of the lattice gauge theory, where the degrees of freedom are the link operators $U_\mu(x)=\operatorname{exp}(i a_\mu A_\mu(x))$ at each lattice point $x \in \Lambda$ with direction $\mu$.
The ultraviolet and infrared divergences in the photon propagator~\cite{PhysRevD.99.034510} are automatically regularised by the finite lattice spacing and volume.

The dynamics of the U(1) gauge theory is described by the standard Wilson action:
\begin{equation}
\label{eq:standard.wilson.action}
 S_W = 
 \frac{\beta}{\xi_0}
 \sum_{x, i}
 \operatorname{Re}
 \left(1 - P_{0i}(x)\right)
 +\beta \xi_0 
 \sum_{x, i>j} 
  \operatorname{Re}
 \left(1 - P_{i j}(x)\right)
 ,
\end{equation}
where $P_{\mu \nu}$ is the plaquette operator:
\begin{equation}
    P_{\mu \nu} = 
    U_\mu(x) \, 
    U_\nu(x+\hat{\mu}) \,
    U^\dagger_\mu(x+\hat{\nu}) \, 
    U^\dagger_\nu(x) 
    \, .
\end{equation}
In the action above, we took into account that $P_{\mu \nu} = 1$ for $\mu=\nu$ and that the trace operator for the $U(1)$ gauge theory is trivial. Moreover, we defined $\beta=2/g^2$, where $g$ is the gauge coupling. 
The parameter $\xi_0$ is the bare anisotropy, which induces an asymmetry between the temporal and spatial parts of the action, measured by the \textit{renormalized anisotropy} $\xi_R=a_t/a_s$. 
We note that for $\xi_0 = 1$ the action is symmetric, which implies $\xi_R=1$.
However, for $\xi_0 = 0$ we \textit{de facto} remove one dimension and hence simulate a different lattice theory.
This implies that the Hamiltonian limit of $\xi_R\to 0$ does not coincide with the naive limit of $\xi_0 \to 0$. Thus, the renormalized anisotropy $\xi_R$ depends on the coupling $\beta$. 
In Secs.~\ref{sec:xiR.potential} and \ref{sec:xiR.glow}, we describe how $\xi_R$ can be determined using the quark static potential and the gradient flow evolution of gauge fields, respectively. 

Instead of using the standard Wilson lattice action in Eq.~\eqref{eq:standard.wilson.action} with the continuum limit $\sim F_{\mu \nu} F_{\mu \nu}$ \cite{peskin2018introduction}, we could have also used a different lattice action with the same continuum limit but other linear combinations of links loops.
Some lattice actions lead to smaller discretization effects~\cite{SYMANZIK1983187, alford1995lattice}, thus reducing the difference between $\xi_R$ and $\xi_0$~\cite{PhysRevD.70.014504}.
Here, however, we intentionally focus on the action in Eq.~\eqref{eq:standard.wilson.action}, because the corresponding Hamiltonian is currently more feasible to simulate.

The results for the observables presented in the following sections have been calculated using the library provided in Ref.~\cite{su2.repo}, 
with which we also generated the gauge configuration samples.
The statistical analysis was done with routines provided in Ref.~\cite{hadron.repo}.

\section{Renormalized anisotropy from the static potential}
\label{sec:xiR.potential} 
At finite lattice spacing, the U(1) gauge theory in 2+1 dimensions is a confining theory~\cite{RevModPhys.51.659, suzuki1983confinement}, with the potential energy $V(r)$ between two static quark charges, $q$ and $\bar{q}$, given by~\cite{POLYAKOV1978477, gopfert1982proof, borgs1983lattice, loan2003path}
\begin{equation}
\label{eq:static.potential.r}
    V(r) = a + \sigma \, r + b \log{(r)}
    \, .
\end{equation}
Here $r$ is the distance separating $q$ and $\bar{q}$, and $\sigma$ is the \textit{string tension}. 
On the lattice, $V(r)$ is found from the large-distance behaviour of the expectation value of Wilson loops~\cite{gattringer2009quantum},
\begin{equation}
\label{eq:Vr.general}
    \lim_{x_\mu \to\infty} 
    \frac{W(x_\mu+ a_\mu \hat{\mu}, r)}{W(x_\mu, r)}
    = \exp(- a_\mu V(r))
    \, ,
\end{equation}
where $\mu$ is any direction on the space-time lattice. 
In the following, we use the conventions $x_\mu = (t,x,y)$ and $\hat{\mu} = (\hat{t}, \hat{x}, \hat{y})$.
We consider planar Wilson loops, where $r$ is the separation between the starting point and the end point of the first half of the loop along a plane perpendicular to the $\mu$ direction.
Which plane we refer to becomes clear from the chosen $r$ coordinate, e.g., the Wilson loop $W(t, x)$ refers to a rectangular loop in the $\hat{t}-\hat{x}$ plane with $r=x$. 

The higher energy states in Eq. \eqref{eq:Vr.general} contribute at finite $x_\mu$, but become exponentially suppressed as $x_\mu$ increases.
Therefore, the potential is found from a fit to a constant of the corresponding effective curve in the large-$x_\mu$ region,
\begin{equation}
\label{eq:Vr.effective}
    a_\mu V^{\text{eff}}(x_\mu, r) = 
    \log \left(
    \frac{W(x_\mu, r)}{W(x_\mu+ a_\mu \hat{\mu}, r)}
    \right).
\end{equation}
On the lattice, we can only compute quantities in lattice units, so the potential in Eq.~\eqref{eq:Vr.effective} is measured in units of $a_\mu$.
On isotropic lattices, we get the same potential for any $\mu$, but on anisotropic lattices, i.e., for $\xi_0 \neq  1$, there is a difference between the \textit{spatial-spatial} and \textit{temporal-spatial} loops, which encodes the renormalized anisotropy $\xi_R$.

In this study, we compute the \textit{temporal} static quark potential with both the \textit{normal}~\cite{PhysRevD.70.014504} and \textit{sideways}~\cite{PhysRevD.63.074501} methods, which we will explain below, see Eqs.~\eqref{eq:wloop.temporal.pot.1} and \eqref{eq:sideways}. We will later use two different approaches to determine the anisotropy $\xi_R$,  see Eqs.~\eqref{eq:normal.potential.fit} and \eqref{eq:matching.sideways}, and will use the normal (sideways) \textit{temporal} quark potential for the first (second) approach. The \textit{spatial} static quark potential is always computed as
\begin{equation}
\label{eq:wloop.spatial.pot}
\lim_{x\to\infty} \frac{W(x+1,y)}{W(x,y)}=\exp(-a_s V(y[a_s])),
\end{equation}
and is used for both ways of determining the anisotropy, i.e., for both Eqs.~\eqref{eq:normal.potential.fit} and \eqref{eq:matching.sideways} below. 
We note that in Eq.~\eqref{eq:wloop.spatial.pot}, $x$ and $y$ are treated on the same footing and thus can be exchanged. 

\begin{figure}[H]
    \centering
    \includegraphics[width=0.95\textwidth]{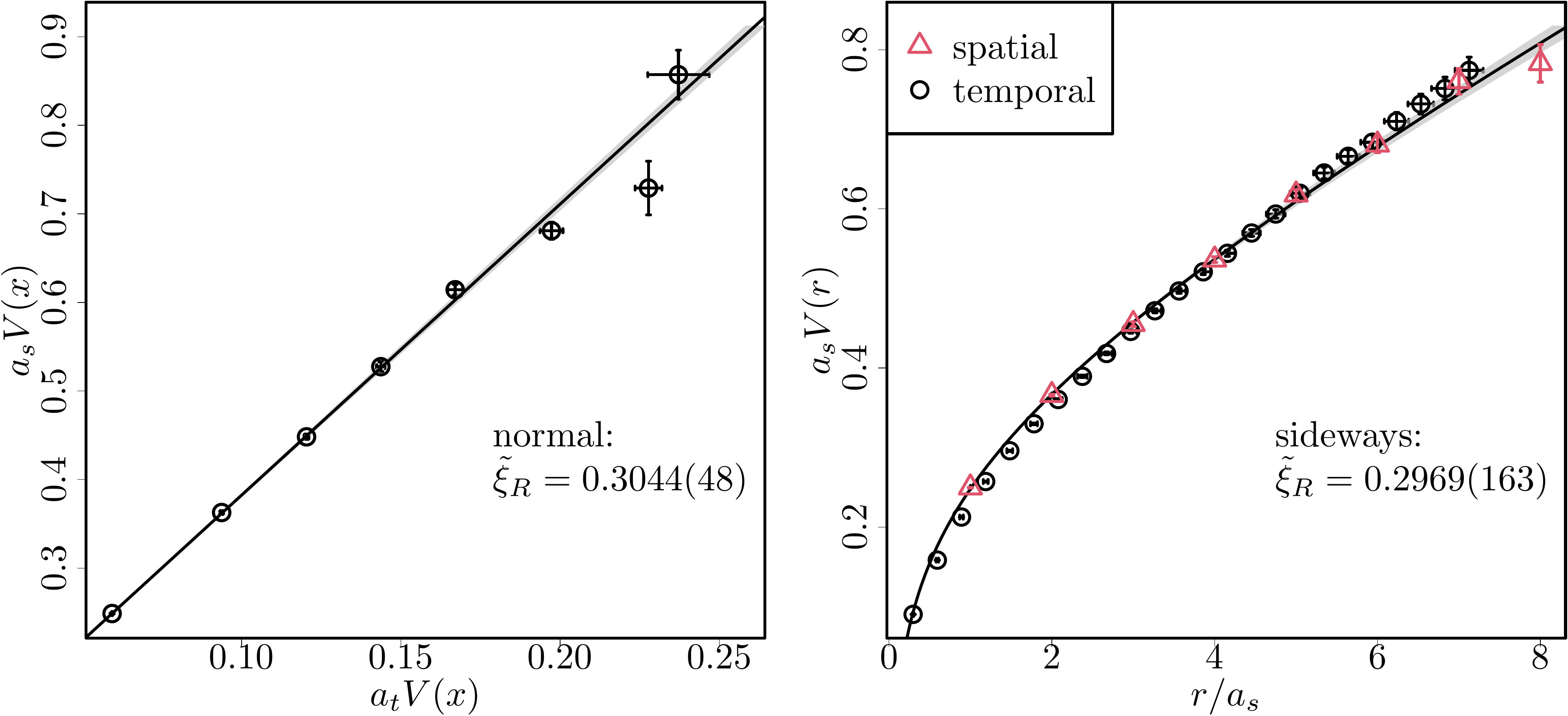}
    \caption{The normal and sideways potentials at $\beta=1.7$, $L/a_s=16$, and $\xi_0={1}/{3}$.
    Left: the normal potentials are fit to Eq.~\eqref{eq:normal.potential.fit}. Right: the sideways potentials are fit to Eq.~\eqref{eq:static.potential.r}.
    The temporal sideways potential is rescaled with $\tilde{\xi}_R$, in order to be in units of $a_s$, with $\tilde{\xi}_R$ as determined from Eq.~\eqref{eq:matching.sideways}.}
    \label{fig:comparisonnormalsidewayspotential}
\end{figure}

\paragraph{Normal potential.}
The normal temporal potential is calculated as
\begin{equation} 
\label{eq:wloop.temporal.pot.1}
\lim_{t\to\infty} \frac{W(x,t+1)}{W(x,t)} = \exp(-a_t V(x[a_s]))
,
\end{equation}
where $a_t$ is the temporal lattice spacing.
In Eqs.~\eqref{eq:wloop.spatial.pot} and \eqref{eq:wloop.temporal.pot.1}, the left-hand sides are equal for $\xi_0=1$ but they differ due to anisotropy effects. Thus, as shown in  Fig.~\ref{fig:comparisonnormalsidewayspotential}(a), we find the renormalized anisotropy $\xi_R$ through a linear fit of $a_s V(x)$ against $a_t V(x)$,
\begin{equation}
\label{eq:normal.potential.fit}
    (a_s V(x)) = 
    \frac{1}{\tilde{\xi}_R} \, (a_t V(x)) \, + \, \tilde{C} 
\, ,
\end{equation}
where $\tilde{C}$ is a constant and $\tilde{\xi}_R$ is an estimator for the renormalized anisotropy $\xi_R$. Here, both $\tilde{\xi}_R$ and $\tilde{C}$ are treated as free parameters of the fit (see the left panel of Fig.~\ref{fig:comparisonnormalsidewayspotential}), and $\tilde{C}$ is the usual arbitrary shift in the potential energy~\cite{griffiths2005introduction}. This method is adapted from the one proposed in Ref.~\cite{PhysRevD.70.014504}.

\paragraph{Sideways potential.} 
Analogously, the sideways temporal potential is determined from
\begin{equation}
  \lim_{x\to\infty}\frac{W(x+1,t)}{W(x,t)}
  =\exp(-a_s V(t[a_t]))
\,  .
\label{eq:sideways}
\end{equation}
Following Ref.~\cite{PhysRevD.63.074501}, the anisotropy is calculated by performing a piecewise linear interpolation of the temporal potential [see Eq.~\eqref{eq:sideways}].
We build the polygonal chain of the temporal potential curve $V(t[a_t])$ as a function of the temporal distance $t/a_t$. 
Then, we restrict ourselves to the points in the linear region, for which $V(r)\approx\sigma r$ [see Eq.~\eqref{eq:static.potential.r}] by requiring $r/a_s \geq 2$. 
For each point of the spatial potential at distance $y/a_s$ in this region, we find the temporal distance $t/a_t$ that matches the temporal and spatial values of the potential on the polygonal chain from the condition:
\begin{equation}
\label{eq:matching.sideways}
V(y[a_s]/\tilde{\xi}_R) = V(t[a_t]).
\end{equation}
The $\tilde{\xi}_R$ needed to fulfill this equation is taken as the estimator for the renormalized anisotropy $\xi_R$.
The final result for $\tilde{\xi}_R$ is computed by averaging over all points  (see the right panel of Fig.~\ref{fig:comparisonnormalsidewayspotential}).

The two methods using Eqs.~\eqref{eq:normal.potential.fit} and \eqref{eq:matching.sideways}, respectively, are summarised and compared in Fig.~\ref{fig:comparisonnormalsidewayspotential},
showing good agreement for the resulting anisotropies $\tilde{\xi}_R$. 
In the remaining part of this paper, we will omit the tilde for brevity and call the numerically determined anisotropies $\xi_R$.

\section{Hamiltonian limit of the plaquette expectation value}
\label{sec:limit.plaquette}
\begin{figure}[!htb]
    \centering
    \includegraphics[width=0.95\textwidth]{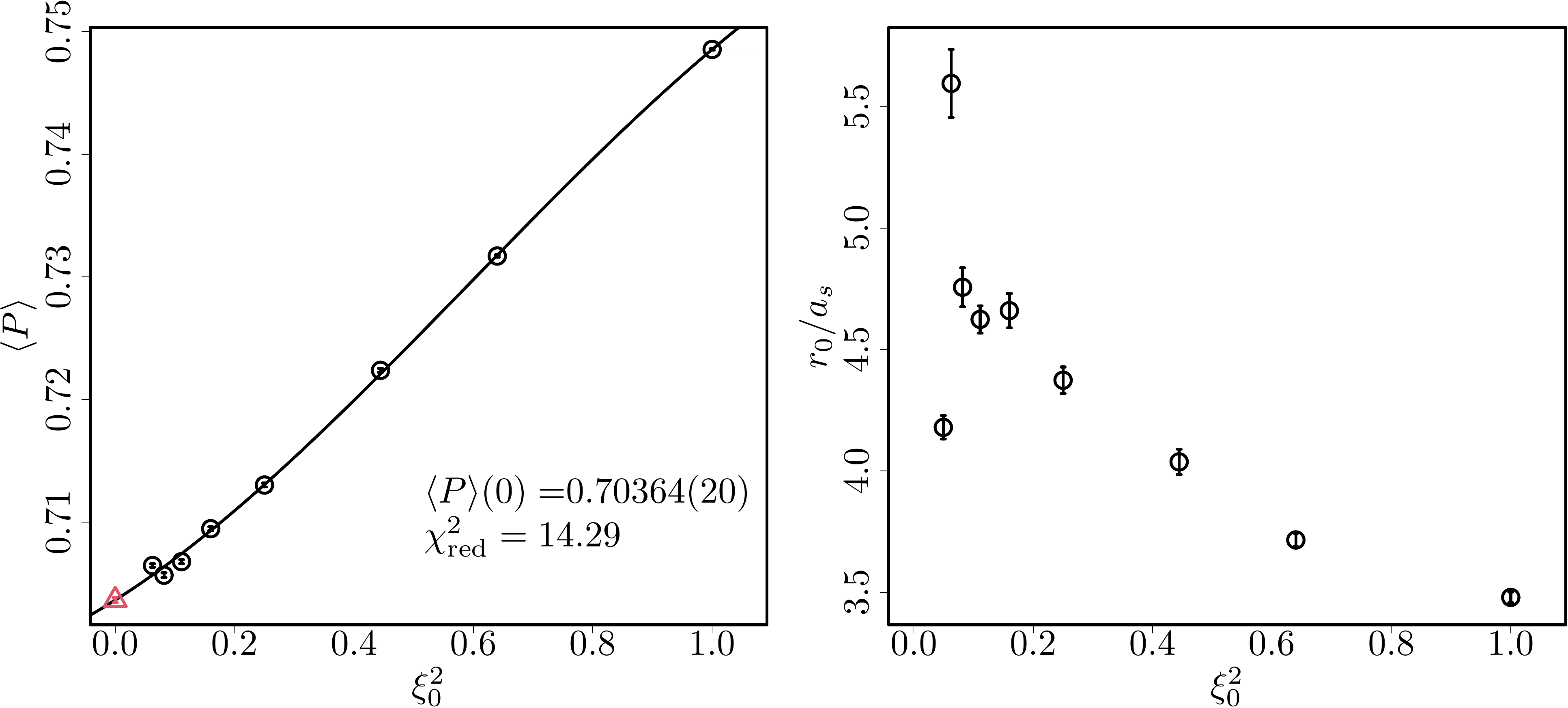}
    \caption{
    Left: naive Hamiltonian limit $\xi_0 \to 0$ of the plaquette expectation value with keeping $\beta$ constant, using $\beta=1.7$ and $L=16$. The plaquette behaves non-monotonously at the smallest anisotropies, which results in a large $\chi^2_\text{red}$ of the cubic fit against $\xi_\text{0}^2$. 
    Right: Sommer parameter in lattice units, $r_0/a_s$, is not constant as a function of $\xi_0^2$, explaining why the naive limit of $\xi_0\to 0$ keeping $\beta$ constant is not sufficient.
    }
\label{fig:naivepotential1.7}
\end{figure}

The naive Hamiltonian limit is reached by sending $\xi_0 \to 0$, while keeping $\beta$ fixed~(see, e.g., Ref.~\cite{loan2003path}). 
However, this is only approximately correct because $\beta$ renormalizes as well, and the discretization effects grow significantly when approaching $\xi_0 \to 0$.
Figure~\ref{fig:naivepotential1.7}(a) shows the failure of this naive approach for the plaquette expectation value $\braket{P}$.

In order to take the correct Hamiltonian limit, the inverse coupling $\beta$ needs to be adjusted, such that the spatial lattice spacing $a_s$ stays constant. 
While this has been previously investigated on the perturbative level~\cite{Carena_perturbative, ByrnesanisotropeSU3}, in our work we focus on a general non-perturbative procedure, which is independent of the coupling strength.

What one should do is to compute the renormalized anisotropy $\xi_R$, using either the static potential (see Sec.~\ref{sec:xiR.potential}) or the gradient flow (see Sec.~\ref{sec:xiR.glow}), and send $\xi_R \to 0$ while keeping $a_s$ fixed.
We then first define the spatial lattice spacing $a_s$ in terms of the Sommer parameter $r_0$~\cite{Sommer_1994}.
Analogously to QCD, we impose the condition~\cite{Sommer_1994}
\begin{equation}
 \displaystyle -r^2\frac{\mathrm{d}}{\mathrm{d}r}V(r)|_{r=r_0}=c_s=-1.65 .
\end{equation}
The value of $c_s$ is arbitrary, but is usually chosen such that the point $V(r_0)$ lies in the region where $V(r)$ is approximately linear. The resulting value for $r_0$ should be in an intermediate region, because the small-$r$ region is hard to probe due to the lattice regularisation, while the large-$r$ region results in a low signal-to-noise ratio in the Wilson loop correlator of Eq.~\eqref{eq:Vr.effective}.
In QCD, its value is fixed to $c_s=-1.65$ resulting in $r_0=0.5$~fm~\cite{Sommer_1994} using experimental inputs; however, we note that this value worked out well also for our data.

\begin{figure}[!htb]
    \centering
    \includegraphics[width=0.95\textwidth]{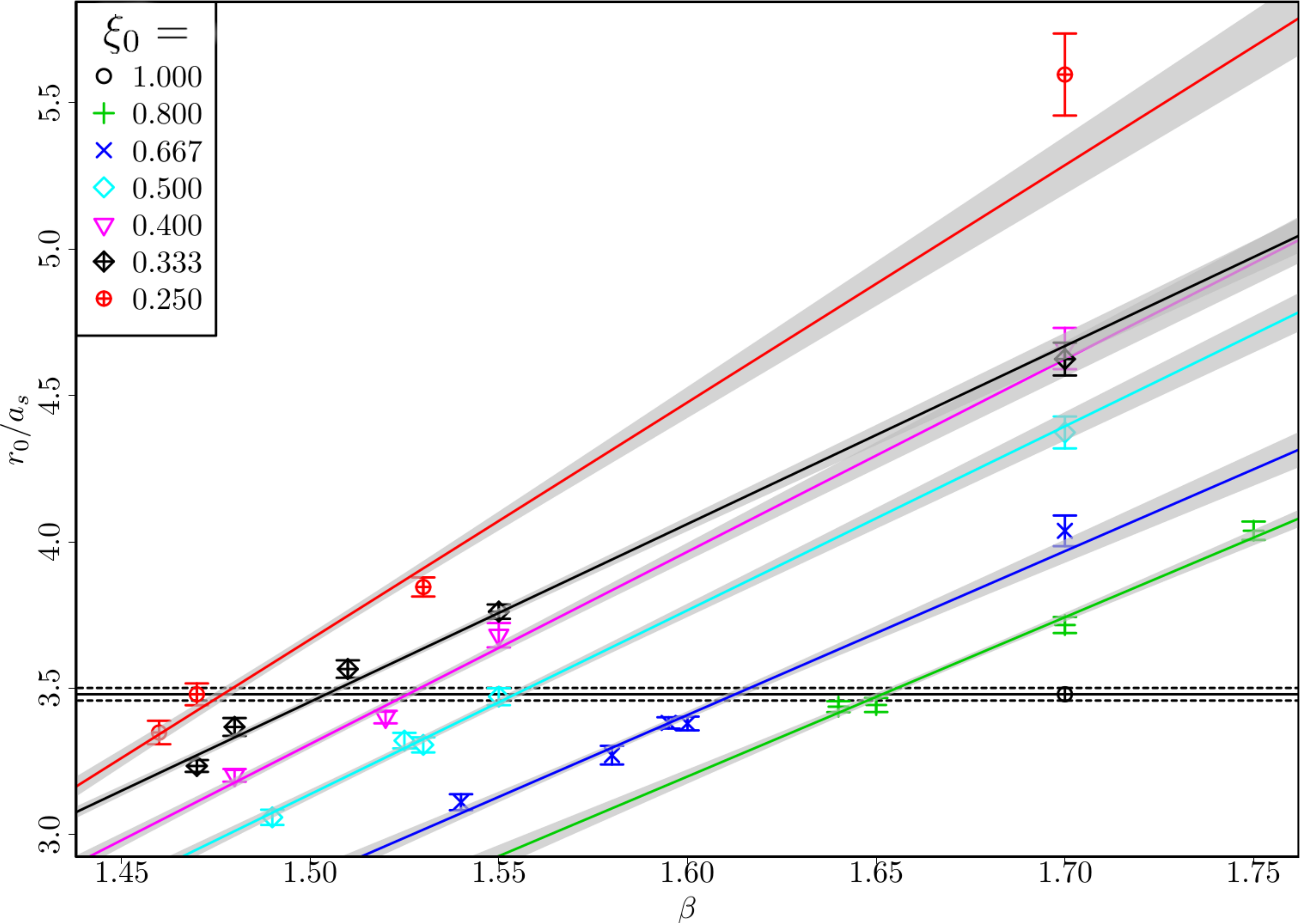}
    \caption{Numerical determination of the coupling $\beta$ such that $\bar{r}_0$ stays constant (black horizontal line). 
    Each marker (see legend) corresponds to a different value of $\xi_0$, and the 
 colored lines are the best-fit interpolations at fixed $\xi_0$. 
    The error bands are shown in faint grey. 
    The intersections of the lines with the horizontal one (given by $\bar{r}_0=\text{const.}$ for $\beta=1.7$ and $\xi_0=1$) determine the values of $\beta$ for which $\bar{r}_0 = \text{const.}$.
    }
    \label{fig:renormr0normal}
\end{figure}

The dimensionless constant $c_s$ implicitly defines $r_0$.
Thus, it is sufficient to keep $r_0/a_s=\bar{r}_0$ fixed throughout the extrapolation of $\xi_R\to 0$.
For the extrapolation, we select a range of values for the bare anisotropy $\xi_0$ and, for each value of $\xi_0$, perform simulations at different values of $\beta$.
Starting from one of these ensembles, we keep $\bar{r}_0$ fixed by interpolating the ensembles with the same $\xi_0$ in the different values of $\beta$ 
(see Fig.~\ref{fig:renormr0normal}). 
To be more precise, we start at $\beta=1.7$ and $\xi_0=1$, compute $\bar{r}_0$, and this value for $\bar{r}_0$ we keep fixed (given by the horizontal line in Fig.~\ref{fig:renormr0normal}). 
We then decrease $\xi_0$, keeping the physical volume approximately fixed. Thus, we increase the number $T/a_t$ of temporal lattice points and keep fixed the number of spatial lattice points, $L/a_s = 16$, such that $\xi_0\equiv L/T$. The physical volume does not change much 
($\xi_0 \approx \xi_R$, see Fig.~\ref{fig:comparisonnormalsidewayspotential}), leading to similar finite-volume effects. 
For a given value of $\xi_0$, we simulate at different bare couplings $\beta$ and fit $\bar{r}_0$ linearly against $\beta$, which gives us the value of the renormalized coupling $\beta_{\text{R}}$, defined through $\bar{r}_0(\beta_\text{R}, \xi_0)=\bar{r}_0(\beta=1.7, \xi_0=1)$ (see the intersecting lines in Fig.~\ref{fig:renormr0normal}). 
Note that we call the coupling $\beta_R$ ``renormalized'' in the sense that is renormalized in $a_t$ but not in $a_s$.
Finally, we interpolate the renormalized anisotropy $\xi_R$ and the plaquette expectation value $\langle P \rangle$ linearly to the values of $\beta$ determined from $\bar{r}_0$.

\begin{figure}[!htb]
    \centering
    \includegraphics[width=0.95\textwidth]{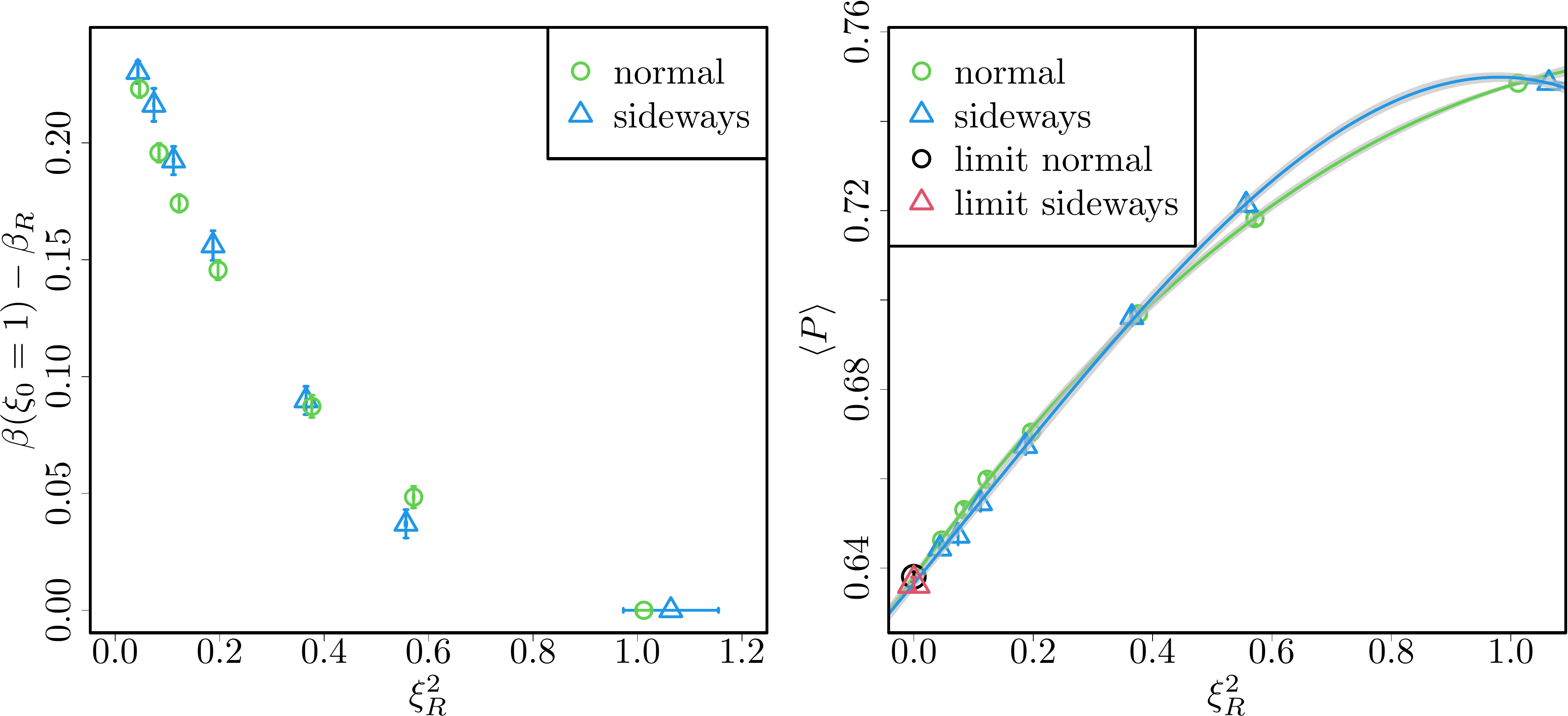}
    \caption{Left: deviation of the renormalized coupling $\beta_R$ from the bare coupling $\beta(\xi_0=1)$, corresponding to the change in $\beta$ required to keep $a_s$ constant.
    Right: correct Hamiltonian limit of $\braket{P}$ using $\beta_R$ instead of keeping $\beta$ fixed. The extrapolation of $\braket{P}$ to $\xi_R\to 0$ is done with a simple cubic ansatz in $\xi_R^2$.}
\label{fig:contlimitnormal}
\end{figure}

The left panel of Fig.~\ref{fig:contlimitnormal} demonstrates how much the renormalized coupling $\beta_R$ deviates from the bare coupling $\beta(\xi_0=1)$, i.e., how much the coupling needs to change in order to keep the spatial lattice spacing $a_s$ constant. In the right panel, we show the resulting correct Hamiltonian limit of $\braket{P}$, which uses the renormalized $\beta_R$ 
 that changes with $\xi_R$, instead of keeping $\beta$ fixed as in the naive Hamiltonian limit (see Fig.~\ref{fig:naivepotential1.7}).
For our initial parameters of $\beta=1.7$ and $\xi_0=1$, the Hamiltonian limit of the plaquette expectation value is $\braket{P}(a_t=0)=0.6381(12)$, as determined with the normal potentials with $\chi^2_\text{red}=0.27$, or similarly $\braket{P}(a_t=0)=0.6365(14)$, as determined with the sideways potentials with $\chi^2_\text{red}=0.34$.

\section{Renormalized anisotropy from the gradient flow}
\label{sec:xiR.glow} 
The renormalized anisotropy $\xi_R$ can be determined using either the static potential (see Sec.~\ref{sec:xiR.potential}) or the gradient flow evolution of gauge fields, as described in the current section.
Using the static potential to find $\xi_R$ has some drawbacks due to the extraction of the signal $V(r)$ from the Wilson loop correlator. 
The temporal extent of the lattice must be large enough, in order to find a plateau for $V(r)$ before the correlator signal degrades~\cite{Lepage:1989hd}. 
Besides this, $a_s V(r)$ increases with decreasing $\beta$~\cite{PhysRevD.11.395}; thus, in the small-$\beta$ region, the leading signal $\exp\{- t V(r)\}$ sooner ends below the statistical noise. 
It is therefore clear that another approach is needed to cover the full parameter space. 

Analogously to QCD~\cite{borsanyi2012anisotropy}, we propose to use the Wilson gradient flow.
First, we evolve in the flow time $\tau$ the gauge links $U_\mu(x, \tau)$ according to the flow equation~\cite{Luscher:2010iy, mazur2017applications}
\begin{equation}
  \label{wflow.eom}
  \frac{d}{d\tau} {U}_{\mu}(x, \tau) =
  -\frac{1}{\beta} \left[\nabla_\mu(x) S_W(U)\right] U_\mu(x, \tau)
  \, ,
\end{equation}
where $\nabla_\mu(x)$ is the covariant group derivative and $S_W$ is the Wilson action defined in Eq.~\eqref{eq:standard.wilson.action}.
Then, we find the flowed ``gauge energies'' as
\begin{equation}
  {E}(\tau) = 
  2 \sum_{x} \sum_{\mu > \nu} 
  \operatorname{Re} \operatorname{Tr} 
  \left[ 1 - P_{\mu \nu}(x, \tau) \right]
  \, .
\end{equation}
At any flow time $\tau > 0$, both perturbation theory and numerical Lattice QCD results unveil that these quantities are already renormalized~\cite{Luscher:2010iy}.
The electric (magnetic) energy $E_{ts}$ ($E_{ss}$) can be obtained by summing over the temporal (spatial)-spatial plaquettes.
Expanding the link operator 
\mbox{$U_\mu(x, \tau) = \exp(i a_\mu A_\mu(x))$}
to $O(a_s^4)$, we find:
\begin{equation}
E_{ts}
\sim a_t^2 a_s^2 \sum_x \sum_{i} F_{0i}^2(x, \tau) 
= a_t^2 a_s^2 V (d-1) \tilde{E}_{ts}  
\, ,
\end{equation}
\begin{equation}
  E_{ss} 
  \sim a_s^4 \sum_x \sum_{i \neq j} F_{ij}^2(x, \tau)
  = a_s^4 V \frac{(d-1) (d-2)}{2} \tilde{E}_{ss}  
  \, ,
\end{equation}
where $V$ is the lattice volume, $d$ is the number of dimensions, and $\tilde{E}_{ts}$ and $\tilde{E}_{ss}$ are the Euclidean average energy densities per space-time direction.
In the continuum limit, there is no distinction among directions, and each pair ${\mu \nu}$ gives the same contribution in physical units.
In that limit, we have $\tilde{E}_{ts}=\tilde{E}_{ss}$; thus, $\xi_R$ can be estimated (up to discretization effects) from the ratio
\begin{equation}
  \zeta(\tau) =
  \sqrt{
  \frac{d-2}{2}
  \frac{E_{ts}(\tau)}{E_{ss}(\tau)}
  }
  \label{eq:zeta}
\end{equation}
at a reference flow time $\tau_0$, given by $\xi_R = \zeta(\tau_0)$.
An advantage of this approach is that a few hundred representative configurations already give a small uncertainty.
The reason is that the gradient flow smooths the fields, suppressing the discretization effects in the flowed plaquette~\cite{Luscher:2010iy}.

\begin{figure}[!htb]
    \centering
    \begin{subfigure}[b]{\textwidth}
    \centering
\includegraphics[scale=0.54]{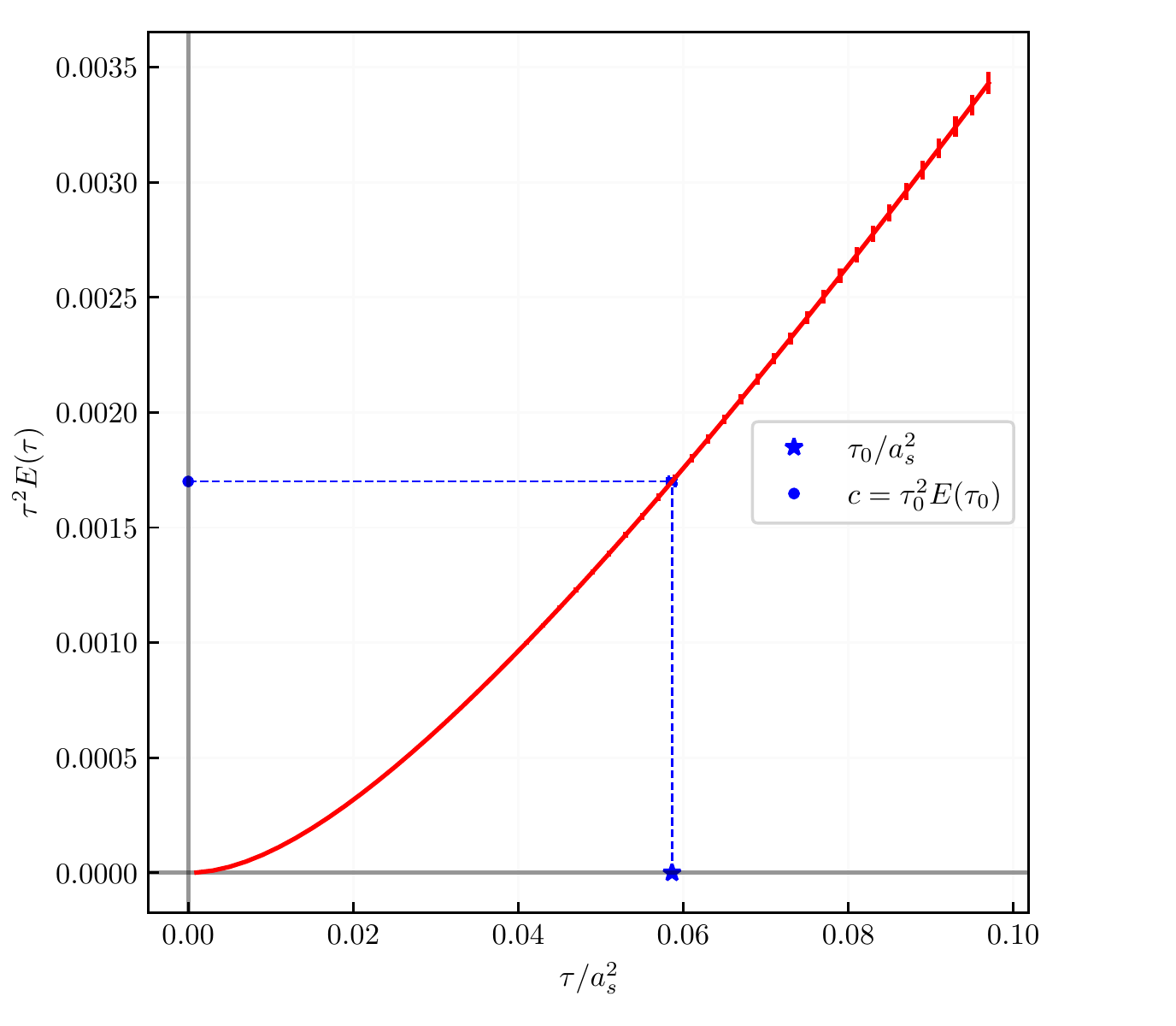}%
        \hfill
    \includegraphics[scale=0.54]{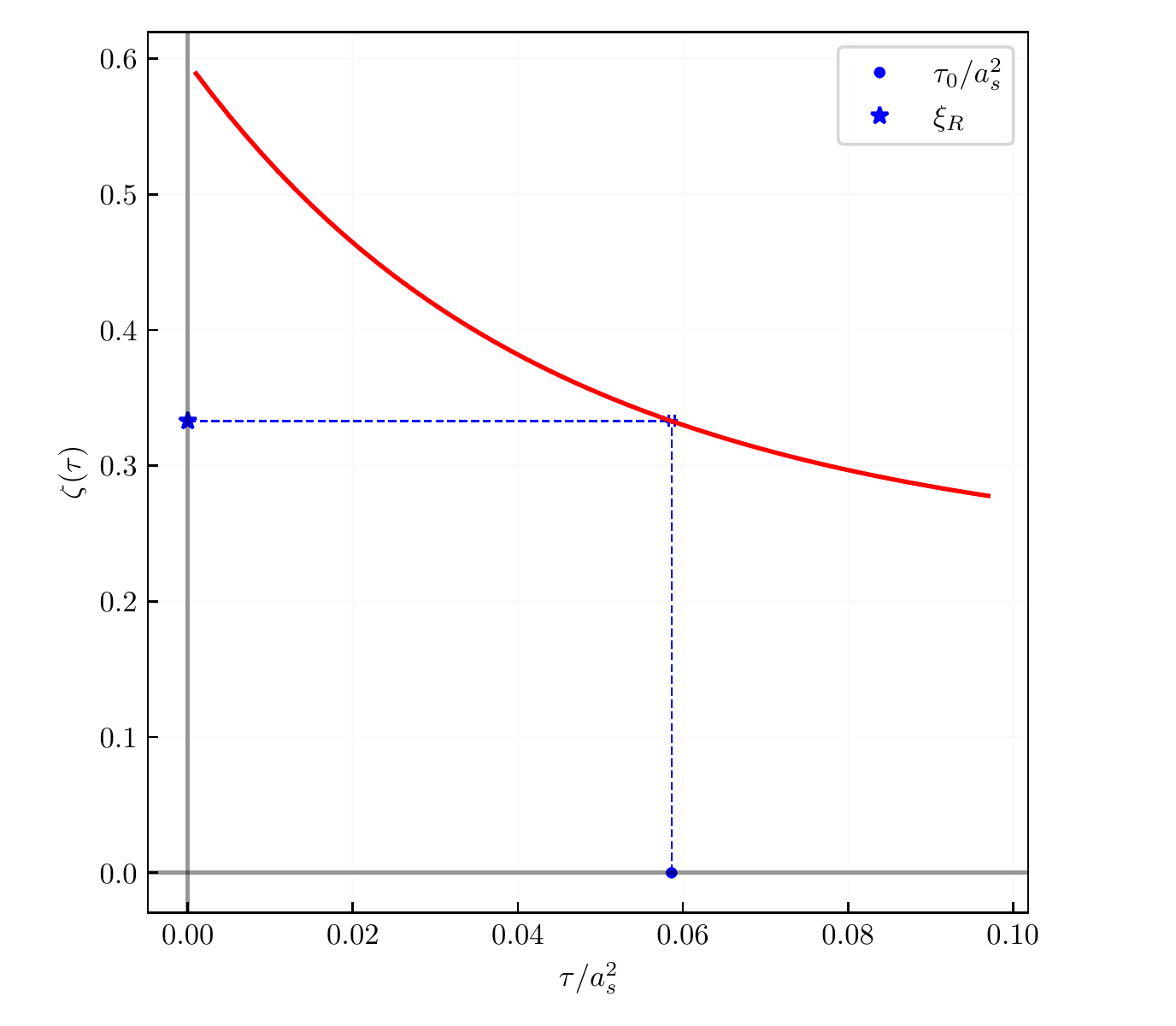}
    \end{subfigure}    
    \caption{
    Left: fixing the reference flow time $\tau_0$ from Eq.~\eqref{eq:gflow.fix.c} (red curve), for the ensemble with $L=16$, $\beta=1.55$, and $\xi_0 = 0.4$. 
    Right: determination of $\xi_R$ using $\tau_0$ from Eq.~\eqref{eq:zeta} (red curve), for the same ensemble.
    Note that both plots contain error bars, which are hard to visualize due to the small uncertainty of the flowed observables (see text).
    }
    \label{fig:tau0.fix.AND.xiR.fix}
\end{figure}

\begin{figure}[!htb]
  \centering
  \includegraphics[scale=0.8]{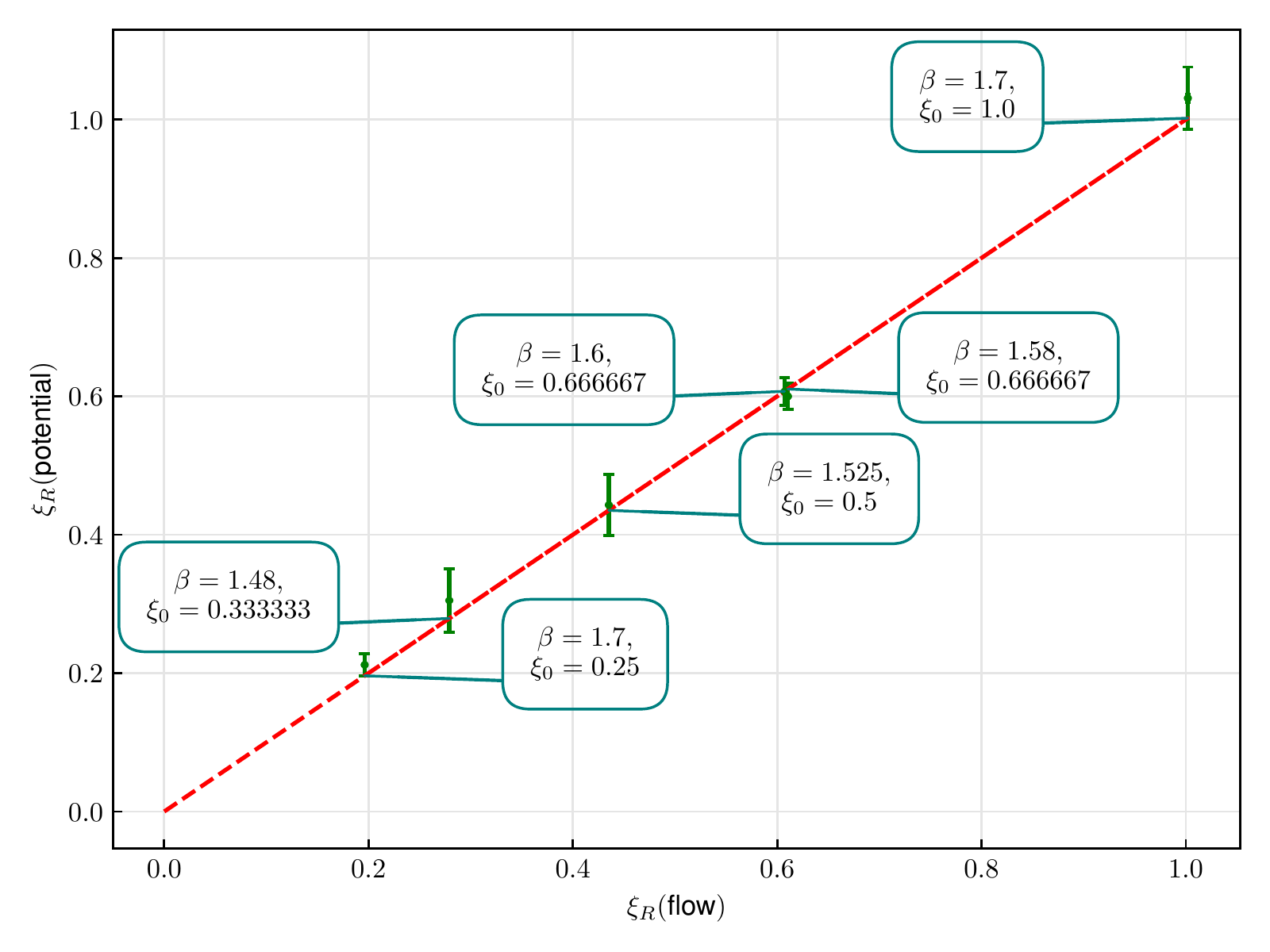}
  \caption{Renormalized anisotropy $\xi_R$ determined with the static sideways potential, compared to $\xi_R$ determined with the Wilson gradient flow using $c=1.628 \cdot 10^{-3}$.
  Each point corresponds to an ensemble, labelled with its value of $\beta$ and $\xi_0$.
  The red line, for which $y=x$, highlights how the choice of a single value for $c$ yields a $1\sigma$ compatibility for all the ensembles. 
   Note that the horizontal error bars are hardly visible due to the small uncertainty of $\xi_R$ determined with the gradient flow (see text).
  }
  \label{fig:xiR.comparison}
\end{figure}

As for $r_0$, the choice of $\tau_0$ is arbitrary, and any positive value would lead to a renormalized quantity.
From perturbation theory, we know that the flowed total energy (in physical units) goes as
${E}(\tau) \sim g^2 \tau^{-d/2}$
\cite{Luscher:2010iy}. We also recall that, in natural units, $[g]=[\text{eV}]^{(4-d)/2}$~\cite{peskin2018introduction} and $[\tau] = [\text{eV}]^{-2}$ from the equations of motion~\eqref{wflow.eom}; 
therefore, we can use the simple condition
\begin{equation}
\label{eq:gflow.fix.c}
    \tau^2 {E}(\tau) \rvert_{\tau=\tau_0} = c
   \, ,
\end{equation}
where 
$c$ is a dimensionless constant.
This procedure is illustrated in Fig.~\ref{fig:tau0.fix.AND.xiR.fix}.
To our knowledge, the value of $c$ for an Abelian $U(1)$ gauge theory in $2+1$ dimensions has not been determined yet, though we mention that in $3+1$ dimensions this was investigated for $SU(N_c)$ theories as a function of $N_c$~\cite{degrand2017simple}.
We determine the value of $c$ such that our results for $\xi_R$ agree with the results for $\xi_R$ obtained from the sideways static potential within $1\sigma$, 
finding $c=1.628(91) \cdot 10^{-3}$.
This result has to be taken with a grain of salt, since it includes discretization effects also due to our prescription for $r_0$.
The uncertainty gives the interval inside which the can vary $c$ without spoiling the compatibility.
In Fig.~\ref{fig:xiR.comparison}, we compare the values found with the two different approaches, using $c=1.628 \cdot 10^{-3}$.

\section{Conclusions and outlook}
\label{sec:conclusions} 

In this work, we have studied the Hamiltonian limit of Lattice QED in 2+1 dimensions.
We have reviewed the issues related to taking this limit, showing an explicit numerical evidence that the naive limit of taking the temporal lattice spacing to zero, $a_t\to 0$, leads to the wrong result.

In particular, we have determined the Hamiltonian limit of the plaquette expectation value $\braket{P}$, along the parameter space trajectory passing through the point $\beta=1.7$ and $\xi=1$.
For this, we have provided a non-perturbative prescription to keep the spatial lattice spacing $a_s$ fixed, while sending the temporal lattice spacing to zero, $a_t \to 0$.
This is done through the renormalized anisotropy $\xi_R= a_t/a_s$, moving along the curve \mbox{$r_0/a_s = \text{const.}\,$}.

We have discussed how to calculate $\xi_R$ using the static quark potential $V(r)$, either the normal or sideways potential.
This approach needs a good signal-to-noise ratio and a low excited-states contamination in the Wilson loop correlators, requiring large enough values of $\beta$ and large enough volumes of the lattice.
Following this observation, we investigated an alternative approach to determine $\xi_R$ using the gradient flow, finding agreement with the first approach for large $\beta$.
In the future, we aim to use this second approach in order to find $\xi_R$ for small values of $\beta$.
This will be particularly relevant for combining quantum computations with classical Monte Carlo computations (as proposed, e.g. in Ref.~\cite{clemente2022strategies}), 
which requires the matching of lattice results obtained in the Hamiltonian and Lagrangian formalisms.

\section*{Acknowledgements}
This work is supported by the Deutsche Forschungsgemeinschaft (DFG,
German Research Foundation) and the NSFC through the funds provided to
the Sino-German Collaborative Research Center CRC 110 “Symmetries
and the Emergence of Structure in QCD” (DFG Project-ID 196253076 -
TRR 110, NSFC Grant No.~12070131001).

L.F.\ is partially supported by the U.S.\ Department of Energy, Office of Science, National Quantum Information Science Research Centers, Co-design Center for Quantum Advantage (C$^2$QA) under contract number DE-SC0012704, by the DOE QuantiSED Consortium under subcontract number 675352, by the National Science Foundation under Cooperative Agreement PHY-2019786 (The NSF AI Institute for Artificial Intelligence and Fundamental Interactions, \url{http://iaifi.org/}), and by the U.S.\ Department of Energy, Office of Science, Office of Nuclear Physics under grant contract numbers DE-SC0011090 and DE-SC0021006.
S.K.\ acknowledges financial support from the Cyprus Research and Innovation Foundation under projects ``Future-proofing Scientific Applications for the Supercomputers of Tomorrow (FAST)'', contract no.\ COMPLEMENTARY/0916/0048, and ``Quantum Computing for Lattice Gauge Theories (QC4LGT)'', contract no. EXCELLENCE/0421/0019.

\bibliographystyle{jhep} 
\bibliography{refs_proceedings_2022}

\end{document}